\documentclass{ws-mpla}

\begin{document}

\markboth{Thomas Sch\"orner-Sadenius}
{The ATLAS Trigger}

%
\catchline{}{}{}{}{}
%

\title{The Trigger of the ATLAS Experiment
}

\author{\footnotesize THOMAS SCH\"ORNER-SADENIUS\footnote{
Now at Hamburg University, Institut f\"ur Experimentalphysik, Luruper Chaussee 149, 22761 Hamburg, Germany.}}

\address{CERN, Division EP\\
1211 Geneva 23, Switzerland\\
schorner@mail.desy.de}

\maketitle

\pub{Received (Day Month Year)}{Revised (Day Month Year)}

\begin{abstract}
With the high bunch-crossing and interaction rates 
and potentially large event sizes
the experiments at the LHC challenge data acquisition and 
trigger systems. Within the ATLAS
experiment, a multi-level trigger system based on hardware and software 
is employed to cope with the task of event-rate reduction. 
This review article gives an overview of the trigger of the ATLAS experiment
highlighting the design principles and the implementation of the system and provides
references to more detailed information.
In addition, first trigger-performance studies and an outlook on the ATLAS 
event-selection strategy are presented.

\keywords{LHC; ATLAS; trigger.}
\end{abstract}

\ccode{PACS Nos.: 07.05.Dz; 07.05.Hd; 07.05.Fb; 07.50.Qx.}

\section{Introduction}

The Large Hadron Collider (LHC), which is currently being built at the 
European Organization for Nuclear Research (CERN) in Geneva\cite{cern}, 
will collide proton 
beams at a center-of-mass energy of 14~TeV and a bunch-crossing rate of 
40~MHz. At the design luminosity 
of ${\rm 10^{34}~cm^{-2}s^{-1}}$ about 
25 proton--proton interactions will take place in every
bunch-crossing. The amount of data that will arise from these conditions
is enormous, making it impossible 
to store the information from all bunch-crossings 
or `events' (note that the ATLAS experiment will have about 
${\rm 10^8}$ electronic channels). Therefore, the experiments at the LHC have to 
provide event selection or `trigger' 
systems that select interesting or even `new' physics processes
and that help reject background processes and known 
(Standard Model) physics processes 
with large cross-sections.

Within the large LHC experimental collaborations (ATLAS, CMS, LHCb), 
the trigger is an important activity. The issues to be addressed range from 
hardware development, through software design and implementation, to the 
development of event-selection criteria. In ATLAS, more than a hundred people
are contributing to these efforts. 

In this review article I will give an 
overview of the trigger system in the ATLAS
experiment\cite{atlas}. After a short general introduction to 
the trigger in Section~\ref{general}, I will turn in more detail to 
the various parts of the multi-level trigger, 
namely the level-1 trigger (Section~\ref{lvl1}) and the
high-level triggers (Section~\ref{hlt}). 
In Section~\ref{performance} I will report on some 
trigger-performance studies. Finally,
Section~\ref{strategy} is devoted to the ATLAS event-selection 
strategy as foreseen for LHC start-up in the  year 2007. 
 
\section{\label{general}The ATLAS Trigger}

\begin{figure}[th]
\centerline{\psfig{file=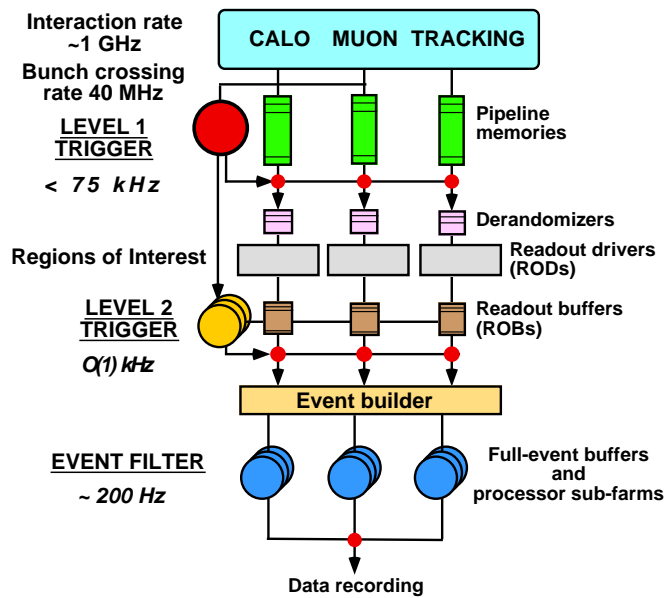,width=3.5in}}
\vspace*{8pt}
\caption{A schematic view of the ATLAS trigger system.}
\label{atlastrigger}
\end{figure}

In the ATLAS experiment, the trigger is designed as a multi-level system
which has to reduce the event rate from 40~MHz to about 200~Hz at which 
events (which will have an average size of the order of 1~MB) can be written to mass 
storage. Figure~\ref{atlastrigger} gives an overview of the trigger system.

The system is divided in three levels (from top to bottom in 
Fig.~\ref{atlastrigger}): 
\begin{itemize}
\item The level-1 (LVL1) trigger is a
  hardware-based system which has to reduce the event rate 
  of 40~MHz to below 75~kHz within a 
  latency\footnote{
    The latency is the time needed to form and distribute the trigger decision. 
    Its maximum value is dictated by the bunch-crossing frequency and the length
    of the pipe-lines in which the event fragments are stored before LVL1 processing.
    } of 2.5~$\mu$s. 
    The LVL1 trigger makes its decision based on 
    comparatively coarse information 
  from only the ATLAS calorimeters and the muon trigger-chamber system.
\item The level-2 (LVL2) trigger, which is part of the high-level trigger
  (HLT), is based on optimized software algorithms running in a processor farm
  and has to reduce the event rate to ${\mathcal O}$(1)~kHz.
  The LVL2 decision, which is based on the result of the LVL1 
  trigger, can take into account the information from 
  all ATLAS subdetector systems which it retrieves as required. The LVL2
  decision has to be ready after about 10~ms.
\item The event filter (EF) is also part of the HLT and is implemented
  using software algorithms. In contrast to LVL2, the EF performs its task 
  only after the complete event has been assembled in the event builder (EB).
  It uses comparatively complex algorithms, based on the offline software, and
  therefore can derive a very detailed event selection and classification, 
  using the best available calibrations. The processing time of the EF 
  for an event is of the order of a few seconds. EF-accepted events are written to 
  mass-storage media.
\end{itemize}

\section{\label{lvl1}The Level-1 Trigger}

\subsection{\label{l1principles}Principles}

\begin{figure}[th]
\centerline{\psfig{file=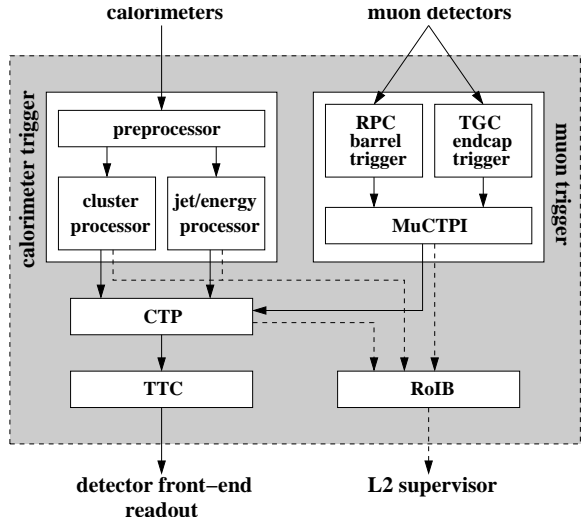,width=3.in}}
\vspace*{8pt}
\caption{A schematic view of the LVL1 trigger. Solid lines between 
  the different components indicate exchange of trigger-object multiplicity
  information, and dashed lines stand for Region-of-Interest. 
  The Region-of-Interest builder
  (RoIB) is not part of the LVL1 trigger but is shown here for 
  completeness. See text for more details.}
\label{l1}
\end{figure}

The task of the LVL1 trigger\cite{l1tdr} is 
to perform a first fast rate reduction, 
while selecting events with interesting signatures in the detector.
Information from the ATLAS calorimeters and from 
dedicated fast muon trigger chambers, the resistive-plate chambers
(RPC) and the thin-gap chambers (TGC), is used for this purpose. 
Consequently, the LVL1 trigger can be viewed in 
three parts, see Fig.~\ref{l1}: 
the calorimeter trigger, which receives the calorimeter 
information and prepares it for the event decision, the muon 
trigger which does the same for the information from 
the muon trigger chambers, and the LVL1 
event-decision part implemented in the central trigger 
processor (CTP). 

The information used to derive the LVL1 event decision is 
given in terms of the multiplicities of physics ``objects" 
detected in the calorimeters or muon-trigger chambers
which have sufficiently high transverse momentum ($p_T$). In the case
of the calorimeter trigger, the objects in question are 
electrons/photons\footnote{At LVL1, electrons and photons 
cannot be distinguished, nor can hadrons and hadronic 
decays of $\tau$ leptons into narrow jets.}, 
$\tau$ leptons/hadrons, and jets. In addition, global energy sums (total 
transverse energy $E_T$, total missing transverse energy $E_{T,miss}$) can be 
considered. 

The LVL1 trigger has to make and distribute 
its decision within a maximum latency of 2.5~$\mu$s. 
During this time, the data fragments of all subdetectors are held in 
pipelined memories from where they are transfered to read-out buffers (ROBs)
upon LVL1 event acceptance (L1A). 
Currently about 1600 ROBs are foreseen in total for the ATLAS experiment. 
The L1A signal is also the starting point 
for the LVL2 trigger which is `seeded' by LVL1 information transfered to it
via the Region-of-Interest builder (RoIB). After the LVL2 decision 
the event fragments in the ROBs are either rejected, or they are 
passed to the event builder.

The TTC system, which is also indicated in Fig.~\ref{l1}, 
has the task of distributing timing and trigger signals
to the read-out electronics. The signals it delivers 
comprise the LHC clock, synchronization signals, the 
L1A signal and test and calibration triggers.
It will not be treated further in this article. 

\subsection{The calorimeter trigger}

The ATLAS calorimeter system\cite{calorimeter} consists of the hadronic iron--scintillator tile
sampling calorimeter in the barrel and
the lead--liquid-argon sampling calorimeters in the barrel and the endcaps.
In addition, the endcaps and the forward directions are equipped with 
hadronic endcap calorimeters with flat copper 
absorbers and copper/tungsten forward calorimeters, respectively. 
These latter two devices also use liquid argon as active medium.

\begin{figure}[th]
\centerline{\psfig{file=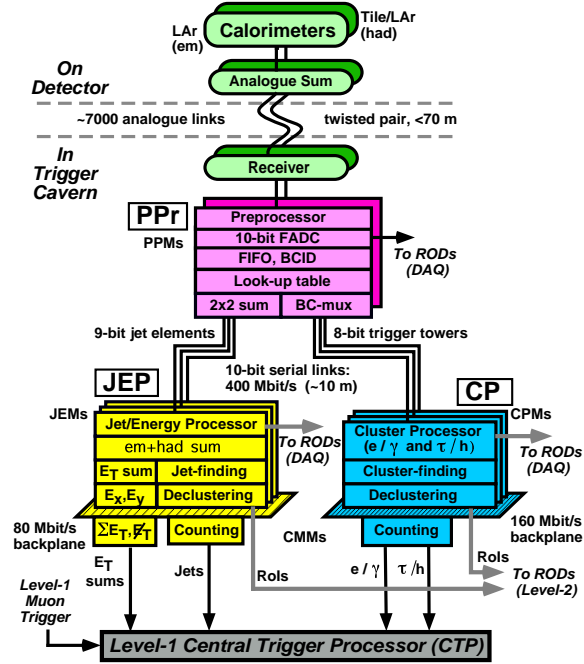,width=3in}}
\vspace*{8pt}
\caption{A schematic view of the calorimeter trigger.}
\label{l1calo}
\end{figure}

The overall architecture of the calorimeter trigger\cite{l1tdr} 
can be seen in Fig.~\ref{l1calo}; it relies heavily on firmware-programmable
FPGAs. On-detector electronics associated with each of the electromagnetic
(EM) and hadronic (HA) calorimeters combines the signals from 
the individual cells by analogue summation.
The results of this combination are 
analogue signals of 7200 approximately projective 
electromagnetic and hadronic trigger towers (TT) with a granularity in 
$\eta\times\phi$ space of 0.1$\times$0.1. 
TTs are arranged so that a HA TT can be found projectively behind each EM TT.

The $\sim$7200 TT signals are transmitted electrically to the preprocessor (PPr) 
electronics\cite{ppr} in the ATLAS electronics cavern where they are digitized in 
fast 10-bit ADCs. The preprocessor also performs bunch-crossing 
identification (BCID) using the pulse shapes of the TT signals, which 
for the LAr calorimeters have a length of several hundred~ns. 
The importance of correct BCID lies in the fact that 
the long (${\mathcal O}$(100)~ns) calorimeter signals have to be associated
to a well-defined bunch-crossing in order to guarantee a sensible
functioning of the trigger. 
After BCID, the PPr uses look-up tables to do a final calibration to 8-bit transverse
energy ($E_T$) values, and presums EM and HA TTs in regions of 
$\Delta \eta \times \Delta \phi = 0.2 \times 0.2$ to so-called jet elements 
which will be used in the jet/energy trigger processor (see later). 

\begin{figure}[th]
  \centerline{\psfig{file=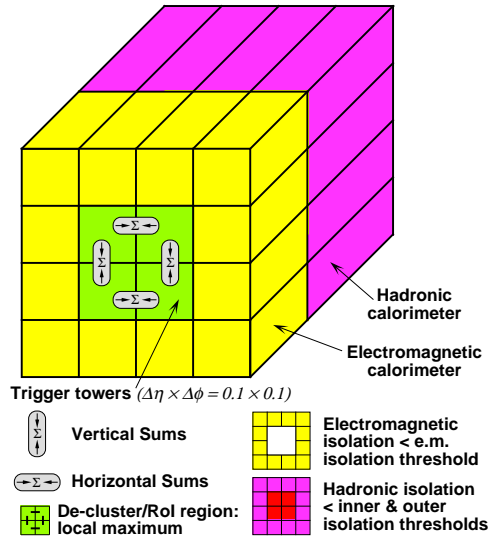,width=2.5in}}
\vspace*{8pt}
\caption{The electron/photon algorithm. See text for details.}
\label{ephoton}
\end{figure}

After the preprocessor, the signal path splits in two. The $\sim$6400 TTs in the 
rapidity range $|\eta| <$~2.5 (corresponding to the inner-detector coverage
and the region of highest EM-calorimeter granularity) are passed to the 
cluster processor (CP). The task of the CP is to identify electron/photon and  
$\tau$/hadron candidates. The algorithm that identifies 
e/$\gamma$ candidates has four elements\cite{alanalg}, see Fig.~\ref{ephoton}:
\begin{enumerate}
\item Clusters of 2$\times$2 EM TTs which are local $E_T$ 
maxima are searched for using a sliding-window technique. These 
2$\times$2 clusters are called Regions-of-Interest (RoIs) and serve as inputs 
to the higher trigger levels. 
\item In a 2$\times$2 cluster there are four pairs of two adjacent TTs. The 
pair with the highest sum $E_T$ defines the transverse energy of the RoI. 
\item The energy in the ring of 12 EM TTs surrounding the 2$\times$2 RoI is 
used to define the EM isolation of the RoI. 
\item Similarly, the 2$\times$2 HA TTs behind the RoI and the 12 HA TTs behind the
EM isolation ring are used to define the HA veto and isolation.  
\end{enumerate}
The $E_T$ of each RoI is compared to one of 8 to 16 thresholds defined in the 
trigger configuration; each RoI passing one of the thresholds contributes to 
the multiplicity count for that threshold (if, in addition, 
the EM and HA isolation variables fulfill criteria associated with the 
threshold in question). 

In the case of the $\tau$/hadron algorithm, the RoI is again of size 2$\times$2, 
but now its $E_T$ is derived from the highest-$E_T$ 2$\times$1 EM TT pair in 
the 2$\times$2 cluster plus the energy of the 2$\times$2 HA TT cluster behind it. 
In addition, 
criteria can be imposed on the electromagnetic and hadronic 12-TT isolation 
rings around the 2$\times$2 core. The $E_T$ of the $\tau$/hadron RoIs
is discriminated against 0--8 programmable thresholds (altogether, there 
exist 16 thresholds of which 8 to 16 may be taken by the e/$\gamma$ trigger; 
only the remaining ones may be used by the $\tau$/hadron trigger).

The results of the cluster processor are 
thus 8 to 16 multiplicities for e/$\gamma$ candidates and 8 to 0 
multiplicities for $\tau$/hadron candidates. These multiplicities
are sent to the central trigger processor (CTP) which makes the 
LVL1 event decision for each bunch-crossing. In addition, for events selected by
LVL1, all selected RoIs (defined by their locations in $\eta$--$\phi$ space and the 
transverse-energy threshold they passed)
are transmitted to the higher trigger levels via the RoIB.

The second signal path from the preprocessor
leads to the jet/energy processor (JEP). In this device, 
candidates for jets are searched for in the matrix of jet elements of 
0.2$\times$0.2 $\eta$--$\phi$ granularity (in the central rapidity range $|\eta| <$~3.2 --
in the forward direction 3.2~$< |\eta| <$~4.9 the algorithm works differently).
This search leads to candidates for (normal) jets and forward jets. Like the 
cluster-processor objects, the jet candidates are located using a 
sliding-window technique, looking for an $E_T$ maximum (RoI) in windows
of 2$\times$2 jet elements. Thresholds are applied for windows of 
2$\times$2, 3$\times$3 or 4$\times$4 elements of $\eta \times \phi$ =0.2$\times$0.2
(window size independently programmable for each threshold).
There are eight programmable jet-$E_T$ and four forward-jet-$E_T$ 
thresholds for which the total 
multiplicities are sent to the CTP. 
The jet RoIs are also sent to the RoIB for events selected by LVL1. 

The jet/energy processor also evaluates the total scalar transverse energy 
and the missing transverse energy of each event, based on all 7200 TT 
signals over the acceptance of $|\eta| <$~4.9. These values, together with 
the total scalar $E_T$ derived from the $E_T$ of all jet RoIs, are also 
discriminated against programmable thresholds, the resulting information 
being passed to the CTP and, for selected events, to the RoIB. 

\subsection{The muon trigger}

The task of the muon trigger\cite{l1tdr} is to find
muon candidates which have transverse momenta in excess of 
one of six programmable thresholds (provided again via the LVL1 trigger menu). 

The muon trigger consists of three separate devices\cite{muonspectro}: 
The RPC trigger prepares the information collected 
in the resistive-plate chamber (RPC) detectors in the ATLAS 
barrel ($|\eta| <$~1.05), the TGC 
trigger does the same for the thin-gap chamber (TGC) information in the 
forward region (1.05~$< |\eta| <$~2.4), 
and the muon-to-CTP interface (MuCTPI) collects information 
from both RPC and TGC triggers, refines it and sends the results to the 
CTP and to the RoIB. 

\begin{figure}[th]
  \centerline{\psfig{file=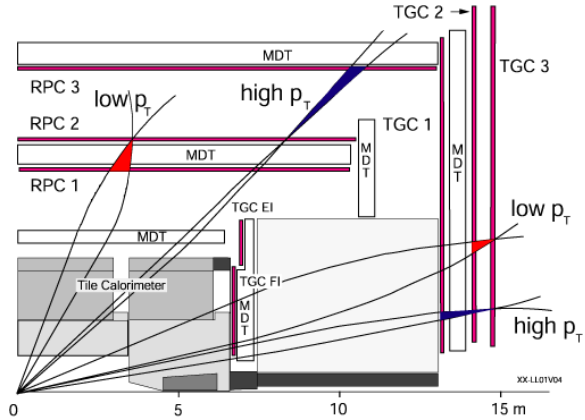,width=3.in}}
\vspace*{8pt}
\caption{The LVL1 muon trigger: An $rz$ view of a quarter of the 
ATLAS detector. Some muon candidate tracks are shown to highlight
the muon trigger algorithm which is explained in the text. `MDT' stands for 
the `Monitored Drift Tubes' precision detectors
which form the muon spectrometer in the barrel.}
\label{muonalgo}
\end{figure}

As can be seen from Fig.~\ref{muonalgo}, there exist three
so-called (RPC or TGC) `stations', each of which contains two planes of 
chambers (the innermost TGC station has three planes).
In the absence of inefficiencies and acceptance gaps, 
this results in six $\eta$ and $\phi$ coordinates for each `view'. 

The algorithm that searches for muon candidates (in the barrel) 
works as follows\cite{l1muonalg}: 
Each hit found in the middle RPC station (RPC2) is extrapolated 
to the innermost RPC station (RPC1) along a straight line through the 
nominal interaction point, and a `coincidence window' is defined around the 
point where this line hits the RPC1 station. Since the ATLAS magnetic field
will deflect charged particles, the size of the coincidence window 
defines the transverse momentum, $p_T$, of muon tracks that can be triggered
upon. A low-$p_T$ muon candidate is found if there is at least one hit in the 
coincidence window and if in at least one of the stations RPC1 and RPC2
hits can be found in both planes and in both views. 
In addition, if there is a coinciding hit in 
at least one of the planes of the outermost station RPC3, a high-$p_T$ 
muon candidate has been found. These low- and high-$p_T$ candidates are the muon trigger
Regions-of-Interest.

Six programmable sets of coincidence windows are defined, each corresponding
to a different $p_T$ threshold; 
three of the thresholds are reserved for low-$p_T$ coincidence
windows (5 to $\sim$10~GeV), and three for high-$p_T$ coincidences ($\sim$10 to 35~GeV). 
Threshold values are designed such that they correspond to an efficiency of 90$\%$.

The muon trigger is arranged in 208 sectors, each 
of which can deliver a maximum of two muon-candidate RoIs to the MuCTPI. 
In case of more than two candidates in one sector, the two 
with the highest $p_T$ values are used and a flag is set. 
The MuCTPI\cite{muctpi} calculates the multiplicity for each $p_T$ threshold, 
applying an algorithm to avoid double-counting of muons, 
and passes the resulting multiplicity values to the CTP for LVL1 event decision.
In addition, for selected events, up to 16 
muon RoIs, defined by their position in $\eta\times\phi$ space
and the transverse-momentum threshold they passed,
are sent to the high-level triggers via the RoIB.

\subsection{LVL1 event decision and LVL1/LVL2 interface}

The LVL1 event decision\cite{l1tdr} is based on the multiplicities of high-$p_T$ objects
sent to the CTP from the calorimeter trigger and the MuCTPI together with threshold
information on global energy sums. The decision 
is derived in two steps. In a first step, the delivered multiplicities are 
discriminated against multiplicity requirements or `conditions', leading to 
truth values `yes' or `no' for each condition defined in the LVL1 trigger 
menu. Then, the condition truth values are logically combined to complex 
`trigger items' which represent signatures to be triggered by LVL1. 
Two examples for such items are 
\begin{center}
`at least two e/$\gamma$ candidates with $p_T >$~10~GeV'\\ 
`at least one jet with $p_T >$~100~GeV and $E_{T,miss} >$~400~GeV'. 
\end{center}
In the first case, the trigger item consists of only one trigger condition;
in the second example, the item consists of the logical `AND' of 
two trigger conditions.
The LVL1 event decision is derived from the 
logical values of all trigger items by 
applying a logical `OR' (see Section~\ref{strategy} for an overview of 
possible signatures).

The final implementation of the CTP is currently being designed. There exists
however a demonstrator prototype, the CTP-D, which has been used
for feasibility studies and tests\cite{ctpd}. The CTP-D receives 32 input signal bits 
encoding the multiplicities of calorimeter/muon trigger objects. 
The multiplicity discrimination is 
implemented using look-up tables, and the
logical combination of conditions to items takes place in programmable 
devices. Up to 32 items can be built. 
The result of the CTP-D includes also prescaling and a 
simple dead-time algorithm for all 32 items. 

In contrast to the CTP-D, the core functionality of the final CTP can probably 
be implemented in one single programmable device, thanks to the speed 
and capacity progress for electronic devices over the past few years. 
160 input bits are foreseen, allowing for a much higher number of multiplicities
to be encoded and thus for a greatly increased trigger flexibility, 
compared to the 32 input bits of the CTP-D. Also the number of trigger 
items will be increased from 32 (CTP-D) to 160 or more.

The LVL2 processing, which will be explained in detail below, starts from
the RoIs selected by the LVL1 trigger. As mentioned above, these RoIs are 
sent to the Region-of-Interest builder\cite{roib} (RoIB) over eight fast links (one for the
muon trigger, six for the calorimeter trigger, and one for the 
CTP information). The RoIB takes all eight data fragments and concatenates them 
into one single data fragment which is transfered to the LVL2 
trigger supervisor assigned for the event. This 
operation must be performed at a full LVL1 output rate of up to 75~kHz without 
introducing significant dead-time into the system. 

\section{\label{hlt}The High-Level Trigger}

\subsection{Overview}

\begin{figure}[th]
  \centerline{\psfig{file=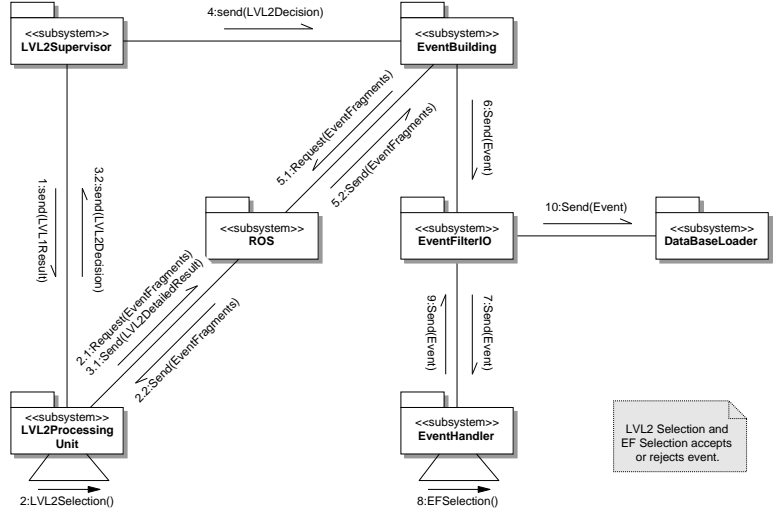,width=4.in}}
\vspace*{8pt}
\caption{Outline of the HLT data flow. See text for more details.}
\label{fighlt}
\end{figure}

The ATLAS high-level trigger\cite{hlttdr} (HLT) consists of 
the level-2 (LVL2) trigger and the event filter (EF)
which are both implemented as pure software triggers running in 
processor farms. Figure~\ref{fighlt} gives an overview of the data flow in the 
HLT environment.

The LVL2 supervisor (L2SV) computers, about ten of which are envisaged 
for the final system, receive the LVL1 result from the RoIB and 
assign events to processors in the LVL2 farms. Note that no processing node 
(not even L2SV nodes) sees the full LVL1 output rate.
In the LVL2 farm processors, 
the LVL2 processing unit (L2PU) is running which forms the interface between 
the L2SV, the read-out subsystem (ROS\footnote{
The ROS aggregates several (typically 3) ROBs (see 
Section~\ref{l1principles}) in a single unit.
}) and the true HLT 
selection software (see Section~\ref{hltssw}).
During the selection procedure,
information from various subdetectors can be retrieved from the ROS. 
In order to minimize the idle-time while waiting for the 
response  to data requests to the ROS, 
LVL2 will allow for multi-threading of selection tasks in the LVL2 processors. 
The decision of the LVL2 selection is sent back to the corresponding 
L2SV which, in the case of a positive LVL2 decision, passes it 
to the event building. LVL2 has a processing time of about 10~ms and has to reduce
the incoming LVL1 rate to ${\mathcal O}$(1)~kHz.

Event building is the data-acquisition step in which all event fragments
from all ATLAS subdetectors are requested from the ROS 
and assembled to give a full ATLAS event. The event is then sent to the 
EF. Here, the EventFilterIO distributes newly arriving events 
to one of the EF processors. 
In these processors, the EventHandler supports the actual EF selection 
(and classification) after which, in case of a positive EF result, the event will be
written to the ATLAS mass-storage devices. The processing time of the EF is limited to 
a few seconds; the EF output rate goal is of the order of 200~Hz.

\begin{figure}[th]
  \centerline{\psfig{file=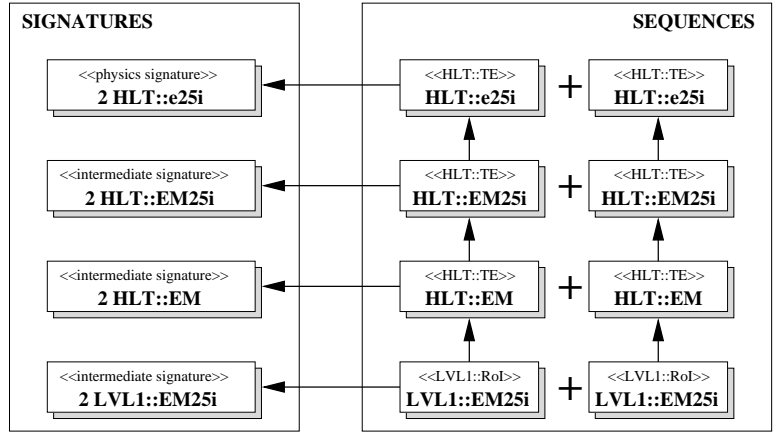,width=4.in}}
\vspace*{8pt}
\caption{Overview on the step-wise selection procedure for an 
  example physics signature. See text for more details.}
\label{stepwise}
\end{figure}

\subsection{HLT Selection Principles}

The LVL2 and the EF share two important principles: 
\begin{itemize}
\item In both levels, the selection procedure starts with a limited 
number of LVL1 (or LVL2, in case of the EF) RoIs which ``seed" the next level.
\item In both levels, the event decision is derived in a step-wise 
procedure in which the initial hypothesis is either confirmed or rejected. 
In each step, the existing information is refined by accessing 
data from more subdetectors or by performing algorithmic work on the present information.
At the end of each step it is checked whether 
the present data still allow for the event to be triggered. If not, the event 
is rejected.
\end{itemize} 
The advantage of these two principles is that they allow for fast 
and light-weight HLT decisions: Working the data-driven way -- i.e. 
starting with only a few RoIs -- significantly 
reduces the amount of data to be moved (at LVL2) and processed (at LVL2 and in the EF)
compared to a scheme where first all data are collected 
and then a global decision based on processing the full event data is performed. 
Only the data necessary for the next step are gathered and/or processed. 
Since an event can be rejected after each of the (small) steps, the amount of time 
invested in rejected events on average is comparatively small.

Figure~\ref{stepwise} highlights the step-wise selection procedure for 
the `2e25i' example physics signature (i.e. for a trigger requirement
of two or more isolated electrons with an $E_T$ of at least 25~GeV): 
In a first step, all existing LVL1 RoIs are collected, 
and it is tested whether they are able to fulfill the 
event signature `2 LVL1::EM25i' which requires two isolated LVL1 
electromagnetic-calorimeter clusters of at least 25~GeV.
If this is the case, algorithms are used to refine the information contained
in the `LVL1::EM25i' RoIs by applying, for example, a shower-shape analysis to the clusters. 
This analysis might be able to distinguish between EM showers due to single electrons
or photons, and EM clusters resulting from $\pi^0 \rightarrow \gamma\gamma$ decays
within jets. Trigger elements
passing this analysis step may thus be regarded as real HLT electromagnetic clusters
and are denoted as `HLT::EM'. If two such trigger elements can be found, the 
next intermediate signature `2 HLT::EM' can be satisfied and 
the selection process will be continued.

In a further step, algorithms might test whether the clusters 
have sufficient transverse energy ($>$~25~GeV) and 
are sufficiently well isolated (suffix `i'), possibly leading to the creation of trigger elements 
`HLT::EM25i' for one or both of the `HLT::EM' trigger elements. In case two 
TEs `HLT::EM25i' are present, the next intermediate signature `2 HLT::EM25i' can 
be satisfied, and the selection process will be continued.
Finally it might be checked whether there are tracks pointing to the calorimeter 
clusters, indicating electrons (that thus can 
be distinguished from photons) and leading to the creation of trigger elements
`HLT::e25i'. If tracks can be found for both `HLT::EM25i' TEs, the physics signature
`2 HLT::e25i' can be satisfied and the event will be triggered.

\subsection{\label{hltssw}HLT Selection Software}

\begin{figure}[th]
  \centerline{\psfig{file=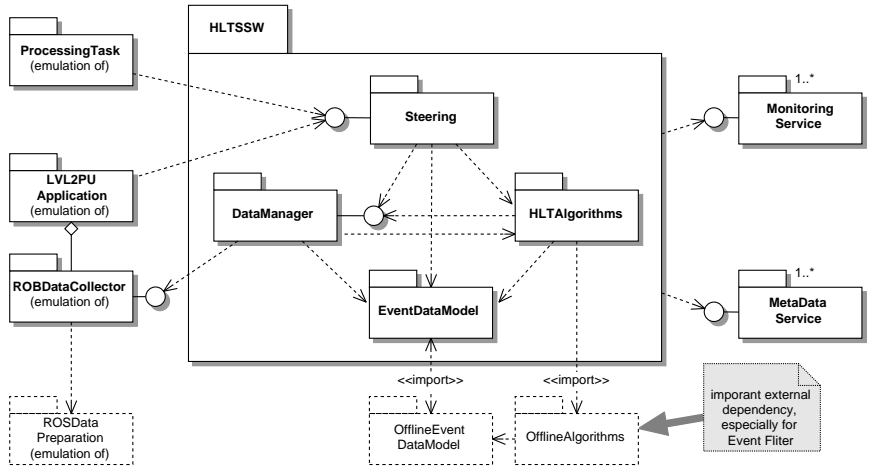,width=4.5in}}
\vspace*{8pt}
\caption{Overview of the HLT selection software.}
\label{hltcore}
\end{figure}

The central part of the HLT clearly are the decision steps performed 
by the selection software running in the 
L2PU and in the EventHandler. This is the HLT selection software 
(HLTSSW). An overview of the core selection software, 
together with its connections and interfaces to other software pieces,
is shown in Fig.~\ref{hltcore}.

The HLTSSW consists of four parts or `packages'\cite{hltarchi}:
\begin{itemize}
\item The Steering\cite{steering} controls the selection software. 
  It organizes the correct order
  of the HLT algorithms processing such that the required data are produced and the 
  trigger decision is obtained. 
\item The event data are structured according to the EventDataModel (EDM). 
  The EDM\cite{edm} covers all data entities in the event and their 
  mutual relations (raw data, LVL1 result 
  and RoIs, LVL2 and EF results). 
\item The HLT Algorithms\cite{hltalg} are used by the Steering to 
  process the event information 
  and to obtain the data on the basis of which the event decision is taken. 
  LVL2 algorithms are special developments, designed for 
  running in the time-critical LVL2
  environment, whereas for the EF selection 
  mostly algorithms adapted for the offline reconstruction will be used.
\item The DataManager is responsible for handling all event data during the 
  trigger-selection procedure and, in particular, 
  for retrieving the necessary additional data from the ROS.
\end{itemize}

The HLT selection software has been developed in the ATLAS offline computing framework
Athena\cite{athena}, which in turn is based on the Gaudi framework\cite{gaudi}. This 
seems natural for the EF which is running offline-reconstruction algorithms, but required
some adaptions of the LVL2 online-software environment. This disadvantage, however, 
has to be compared to the advantages: First, a common HLT framework allows
for a flexible boundary between LVL2 and EF, facilitating latency and rejection 
trade-offs between these two systems. Second, the offline framework is the 
standard user and analysis framework in ATLAS. Thus anybody capable of using this framework
should be able to develop selection and reconstruction algorithms not only for EF, but
also for LVL2.

\begin{figure}[th]
  \centerline{\psfig{file=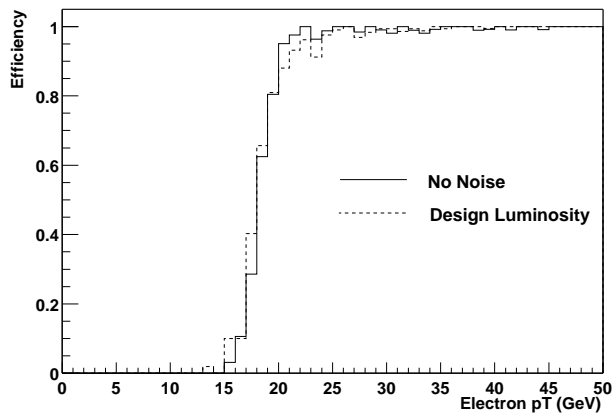,width=8.cm}}
\vspace*{8pt}
\caption{LVL1 efficiency for single electrons as a function of their 
transverse energy, using a 17~GeV threshold for different scenarios. 
See text for more details.}
\label{l1effi}
\end{figure}

\section{\label{performance}Trigger-Performance Studies}

Already some time ago thorough trigger-performance studies were carried 
out in ATLAS; they are documented, for example, in the HLT 
Technical Proposal\cite{tp}. In addition, the recently published 
HLT Technical Design Report\cite{hlttdr}
has initiated a large number of studies which involve improved knowledge compared
to the Technical Proposal and which therefore promise improved insight into the 
ATLAS physics potential. 
These studies range from validations of the 
LVL1 trigger simulation (which provide the input to the 
HLT studies), through feasibility studies for the HLT architecture, to 
stability and rate tests for the selection software. All results shown below are 
taken from the Technical Design Report, if not stated differently.

Figure~\ref{l1effi} shows, as a function of the electron $E_T$, the simulated trigger 
efficiency for electrons using a trigger for single EM-calorimeter objects\cite{edvalid}.
An $E_T$ cut of 17~GeV has been applied on the raw measured energy in order
to achieve an efficiency of 95~$\%$ for the nominal threshold of 20~GeV.
A sharp rise of the efficiency around the
nominal threshold value can be observed. The behavior is similar for two
scenarios considered, namely the pure signal without calorimeter noise or pile-up 
added (label `no noise')
and high (LHC design) luminosity (${\rm 10^{34} cm^{-2} s^{-1}}$, labeled `Design Luminosity'). 

\begin{figure}[th]
  \centerline{\psfig{file=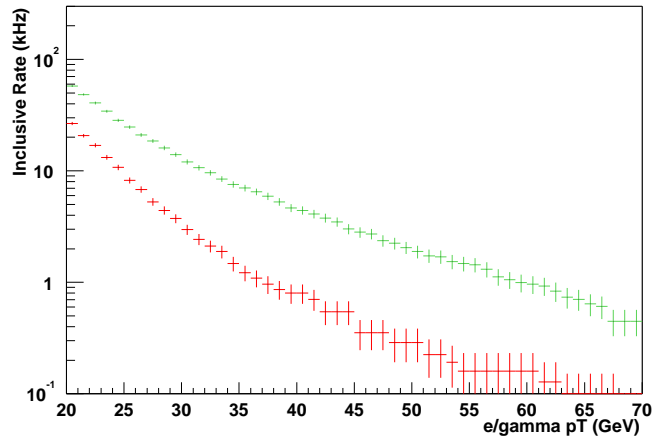,width=8.5cm}}
\vspace*{8pt}
\caption{LVL1 rate for the inclusive EM-cluster trigger 
versus $E_T$ threshold without (top) and with (bottom) isolation requirements 
at low luminosity 2$\cdot {\rm 10^{33} cm^{-2} s^{-1}}$.
See text for more details.}
\label{l1rate}
\end{figure}

Figure~\ref{l1rate} shows, as a function of the $E_T$ threshold, 
the expected trigger rates for a LVL1 calorimeter
trigger for single electrons for 2$\cdot {\rm 10^{33} cm^{-2} s^{-1}}$.
The $E_T$ threshold scale is defined such that the efficiency for genuine 
electrons with $E_T$ equal to the cited value is 95~$\%$. 
The top line indicates the rate for the
case of no isolation required; the bottom line shows the expected rate
with isolation cuts applied on the electromagnetic calorimeter cluster.
The difference between the two lines demonstrates the ability of the isolation
criteria to reduce the dominating rate contribution from misidentified jets.

Similar studies are currently being performed for the LVL1 muon trigger.

Table~\ref{tabl2calo} shows an example of a HLT study taken mainly from the 
Technical Proposal\cite{tp} in which the 
expected performance of the isolated-electron HLT trigger was tested\cite{tpem}. 
Shown are, 
separately for low and design luminosity, the expected trigger rates and efficiencies 
and first timing measurements\footnote{
Note that at the time of the Technical Proposal, the low luminosity scenario -- in 
contrast to today's assumption of 2$\cdot {\rm 10^{33} cm^{-2} s^{-1}}$ -- was
assuming ${\rm 10^{33} cm^{-2} s^{-1}}$.
}. 
In this table, the rates, efficiencies $\epsilon$ and timings are shown for the various steps
of the HLT electron-trigger process: `LVL2 Calo' corresponds to the precision 
reconstruction of the EM calorimeter cluster, and to the measurement 
of the transverse energy of the cluster. `LVL2 Precision' and `LVL2 TRT' refer to 
two different track-finding algorithms involving 
the precision silicon and pixel detectors
and the ATLAS transition-radiation tracker (TRT), 
respectively. `LVL2 Matching' denotes
the effort to match the track and cluster information in position and energy. 
`EF Calo (Matching)' are the EF equivalents to `LVL2 
Calo (Matching)', and `EF ID' stands
for RoI-seeded track search in the ATLAS inner (tracking) detector (ID). 
Also shown in the table (column 2$\cdot{\rm 10^{33}}$) are first rate studies from the 
HLT Technical Design Report for some of the steps mentioned above\cite{edvalid,bainesnew}; 
the efficiencies are expected to be between the values for ${\rm 10^{33} cm^{-2} s^{-1}}$ and 
 ${\rm 10^{34} cm^{-2} s^{-1}}$.

\begin{table}[th]
\tbl{Overview of HLT trigger rates, efficiencies and timings for the single electrons
at different luminosities. See text for more explanation.}
{\begin{tabular}{@{}c|ccc|ccc|c@{}} \toprule
Lumi [${\rm cm^{-2}s^{-1}}$] & & ${\rm 10^{34}}$ & & & ${\rm 10^{33}}$ & & 2$\cdot{\rm 10^{33}}$\\ \colrule
Trigger & Rate & $\epsilon$ & Timing & Rate & $\epsilon$ & Timing & Rate\\ 
 Step   & [Hz] & [$\%$]     &        & [Hz] & [$\%$]  &  & [Hz] \\ \colrule
LVL1 output & 21700 & 94.6 & -- & 11000 & 92.6 & -- & 12000 \\ \colrule
LVL2 Calo & 3490 & 97 & 0.3~ms & 1100 & 96 & 0.2~ms & 2114 \\
LVL2 Precision & 620 & 90 & 13~ms & 150 & 92 & 6~ms & -- \\
LVL2 TRT & 1360 & 90 & 1.2~s & 360 & 89 & 210~ms & -- \\
LVL2 Matching & 460 & 85 & -- & 140 & 88 & -- & 137 \\ \colrule
EF Calo & 313 & 84 & 0.63~s & 85 & 86 & 0.56~s & -- \\
EF ID & 149 & 79 & 71~s & 57 & 82 & 1.6~s & -- \\
EF Matching & 117 & 78 & -- & 41 & 81 & -- & 30 \\
\end{tabular}}
\label{tabl2calo}
\end{table}

The efficiencies are shown for electrons of 20/25/30~GeV transverse energy for
${\rm 10^{33}}$/2$\cdot{\rm 10^{33}}$/${\rm 10^{34} cm^{-2}s^{-1} }$, respectively,
over the calorimeter rapidity range of $|\eta| <$~2.5. 
The timing measurements, which were performed on various computing platforms with 
the results being transformed to a 500~MHz Pentium PC equivalent (year 2000),
give the latency within which 95~$\%$ of all 
events were processed; median numbers are in some cases much shorter. The numbers were derived 
for the purely algorithmic part of the trigger process, 
excluding as much as possible input/output processes and data preparation.
Efficiency and rate values for the HLT are given with respect to the LVL1 efficiency 
of about 95~$\%$ and the LVL1 output rates which are also given in the table.
The final rates for the two lower-luminosity scenarios are in acceptable 
agreement with the foreseen trigger menu, see Section~\ref{strategy}.

Table~\ref{hltmuon} shows
the estimated output rates of the LVL2
muon trigger algorithm $\mu$FAST\cite{hlttdr,mufast} for various physics processes, 
applying a $p_T$ threshold of 6 or 20~GeV\footnote{
Note that according to the present trigger configuration ideas presented 
in Section~\ref{strategy} no inclusive 6~GeV--muon trigger is foreseen anymore 
for the low luminosity scenario.
} for the ${\rm 10^{33}cm^{-2}s^{-1}}$ or 
${\rm 10^{34}cm^{-2}s^{-1}}$ luminosity scenario, respectively. For this study
the rapidity of the muons was limited to the barrel region $|\eta| <$~1. The $\mu$FAST 
algorithm was especially developed for the LVL2-online environment and relies on information 
from both the muon trigger chambers (RPCs) and the muon precision chambers (MDTs). It 
is seeded by LVL1 RPC RoIs which help to define a `road' around the $\mu$ trajectory. The 
MDT tubes that are hit by such a road are selected, and a straight-line fit is 
placed through all selected tubes from which an estimate of the $\mu$ $p_T$ can be derived.

\begin{table}[th]
\tbl{Rates for the LVL2 muon trigger algorithm $\mu$FAST with a 
threshold of 6~(20)~GeV for low(high) luminosity for various physics processes.}
{\begin{tabular}{@{}c|cc@{}} \toprule
Physics    & ${\rm 10^{33}cm^{-2}s^{-1}}$  & ${\rm 10^{34}cm^{-2}s^{-1}}$  \\
 Process   & [kHz] & [kHz]     \\ \colrule
$\pi$/K decays & 3.00 & 0.07 \\
b decays & 0.90 & 0.09 \\
c decays & 0.50 & 0.04 \\
W$\rightarrow\mu\nu$ & negligible & 0.02 \\
cavern background & negligible & negligible \\ \colrule
Total & 4.40 & 0.22 \\
\end{tabular}}
\label{hltmuon}
\end{table}

\begin{figure}[th]
  \centerline{\psfig{file=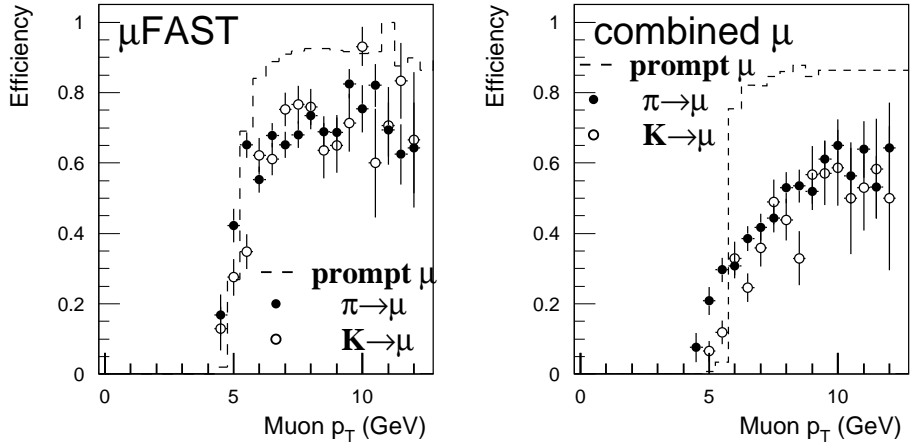,width=12.cm}}
\vspace*{8pt}
\caption{The efficiency at LVL2 with respect to LVL1 of the $\mu$FAST and the combined muon
algorithm for prompt muons and muons from K/$\pi$ decays in flight.
See text for more details.}
\label{mufastps}
\end{figure}

Figure~\ref{mufastps} shows, as a further result from the new HLT studies, the LVL2 efficiency 
for prompt muons and for K/$\pi$ decays in flight. The left plot shows the efficiency for 
the $\mu$FAST algorithm; the efficiency curves were derived with a more complete `combined'
muon algorithm which combines information from the ATLAS tracking detectors and
the muon spectrometer. It is clearly visible that the combined algorithm provides increased
separation power between prompt muons and muons from K/$\pi$ decays which in turn leads to a
decreased rate for muons with $p_T >$~6~GeV of 2.1~kHz at ${\rm 10^{33}cm^{-2}s^{-1}}$.

Further studies, for example those investigating technical details of the HLT architecture,
cannot be discussed here. Please refer to some recent publications on these issues\cite{recent}.

\section{\label{strategy}ATLAS Event-Selection Strategy}

The aim of the ATLAS event selection is to be as open and 
efficient as possible for new physics processes, while preserving good rejection power 
against well-known background and Standard Model processes with large cross-sections\cite{dettdr}.
This aim shall be achieved by designing a selection based on low 
multiplicities of high-$p_T$ objects. It is foreseen to implement the 
following triggers\footnote{
Here only the unprescaled part of the trigger menu is mentioned. There 
are numerous other (prescaled) triggers which will be used for more 
exclusive selections, for example in the area of B physics, or for 
monitoring, calibration and efficiency-determination purposes.
}:
\begin{itemize}
\item Inclusive and di-lepton triggers will cover large parts of the 
  Standard Model and the discovery physics programme of ATLAS. Examples of
  processes accessible with them are associated 
  Higgs production ${\rm t{\overline t}H}$, H$\rightarrow$ZZ(WW) decays,
  top physics, or Z$\rightarrow {\rm l{\overline l}}$ decays
  (also needed for detector calibration purposes). The B-physics 
  programme relies to a large extent on muon triggers augmented by
  more exclusive selections.
\item There will be a variety of jet triggers, with required 
  multiplicities between one and four jets. The thresholds will be 
  highest for the inclusive jet trigger, of the order of 400~GeV (on HLT); 
  for the four-jet trigger, thresholds of around 100~GeV are envisaged. In 
  addition, jets in the very forward direction, close to the proton beams, 
  can be used to trigger. The purpose of the jet triggers is mainly 
  QCD studies (and background determination of search channels), but also new 
  physics signatures can be selected with them 
  (for example new resonances with the di-jet trigger or R-parity violating supersymmetry). 
\item Triggers based on the missing transverse energy and the total 
  transverse energy are vital parts of the search for new physics, 
  for example for invisibly-decaying particles and supersymmetric signatures. 
  The threshold for the 
  missing-$E_T$ trigger is assumed to be about 150~GeV, and the total-$E_T$ 
  trigger should fire at energies above 1~TeV approximately. 
\item In addition to the above, there are a number of `mixed' triggers 
  foreseen, in which leptons and jets are combined with 
  missing transverse energy, for example.
\end{itemize}

The threshold values of the various triggers mentioned above are 
mostly the result of physics studies involving 
leading-order Monte Carlo predictions
and (incomplete) simulations of the ATLAS detector and its trigger. 
They represent compromises between selection efficiency and rate reduction needs
and might well be subject to future changes, due to 
more precise theoretical predictions, better ATLAS detector simulations and 
a better knowledge of the ATLAS detector layout.

All in all, it is foreseen that ATLAS will trigger on a set of about 100
physics signatures, including prescaled triggers and calibration 
and monitoring triggers. The total rates for the LVL1 trigger and the HLT that
are currently aimed for are 25~kHz and 200~Hz, respectively, taking into consideration
various uncertainties on the predicted rates and the reduced rate ability ATLAS has
to cope with during the start-up phase due to financial problems. 
Table~\ref{ratetab} gives estimates of rates for 
the most important unprescaled trigger signatures\cite{trigconfig} for the low
luminosity scenario (${\rm 2\cdot 10^{33}cm^{-2}s^{-1}}$). The numbers quoted in 
this table are mainly derived from older predictions (for example from the 
Technical Proposal\cite{tp}) for the original low-luminosity scenario
${\rm 10^{33}cm^{-2}s^{-1}}$ which were scaled appropriately.

\begin{table}[th]
\tbl{Rate estimates for various 
trigger signatures at LVL1 and HLT for the low-luminosity scenario
${\rm 2\cdot 10^{33}cm^{-2}s^{-1}}$ . 
`EM', `MU' and `J' stand for 
LVL1 electron/photon, muon or jet candidates, respectively. `xE' denotes
missing transverse energy at both LVL1 and HLT. 
`e', `$\gamma$', `$\mu$' and `j' denote electron, photon, muon and 
jet candidates at HLT. Only an extract from the complete inclusive 
unprescaled trigger menu is shown here; $\tau$/hadron triggers, forward-jet
triggers and pure energy triggers are completely omitted. The `2MU6' threshold
is under discussion; a threshold value as low as possibly allowed by the 
muon-trigger design will be used. The `mass' that is referred to in the `2MU6' line
is the invariant mass of the di-muon system which may be required to be close to the mass of
the J$\backslash\Psi$, for example, from the decay of which 
the muons are supposed to come. }
{\begin{tabular}{@{}ccccc@{}} \toprule
LVL1 & LVL1 Rate & HLT & HLT Rate & Purpose\\ 
  Signature& [kHz]   & Signature & [Hz] & \\ \colrule
MU20 & 0.8 & $\mu$20i & 40 & ${\rm t{\overline t}}$H, H$\rightarrow$WW,ZZ,\\
     &     &          &    & top, W', Z', Z$\rightarrow$ll \\
2MU6  & 0.2 & 2$\mu$10, 2$\mu$6+mass & 10,10 & H$\rightarrow$WW,ZZ,\\
     &     &          &    & B, Z$\rightarrow$ll \\
EM25i & 12 & e25i,$\gamma$60i & 40,25 & ${\rm t{\overline t}}$H, H$\rightarrow$WW,$\gamma \gamma$\\
     &     &          &    & top, W', Z', Z$\rightarrow$ll, W$\rightarrow\nu$l\\
2EM15i & 4 & 2e15i,2$\gamma$20i & $<$1,2 & H$\rightarrow$WW,ZZ,$\gamma \gamma$\\
       & & & & Z$\rightarrow$ll\\
J200 & 0.2 & j400 & 10 & QCD, new physics\\
3J90 & 0.2 & 3j165 & 10 & QCD, new physics \\
4J65 & 0.2 & 4j110 & 10 & QCD, new physics \\
J60+xE60 & 0.4 & j70+xE70 & 20 & Supersymmetry \\
MU10+EM15i & 0.1 & $\mu$10+e15i & 1 & H$\rightarrow$WW,ZZ \\
          &     &               &  & ${\rm t{\overline t}}$ fully leptonic \\
\end{tabular}}
\label{ratetab}
\end{table}

\section{Conclusion and Outlook}

The ATLAS experiment will run in the harsh environment of the LHC, with 
a high bunch-crossing frequency, large event 
size and high luminosity. In order to 
select interesting physics processes from the 
bulk of background and known physics
processes, a multi-level trigger system has 
been designed, consisting of a hardware 
LVL1 trigger and the software levels LVL2 and EF (the HLT). 

Almost all parts of the LVL1 trigger system are designed 
or even already built. For the HLT, the recently published 
Technical Design Report\cite{hlttdr} marks an important step towards the 
final design. In addition, it has initiated a 
new round of trigger-performance studies.
In parallel to the hardware and software activities connected to the 
development of the LVL1 trigger and the HLT, detailed studies of 
possible trigger menus for data taking in the LHC 
start-up phase are currently being performed in close collaboration with 
the ATLAS physics working groups.

All in all, the ATLAS trigger is on a promising path, 
clearly aiming for meeting 
all requirements necessary for smooth ATLAS data taking from 2007 onwards.

\section*{Acknowledgments}

I would like to thank N.~Ellis, S.~Tapprogge, E.~Eisenhandler and 
L.~Nisati for their careful reading of the manuscript and all 
my ATLAS colleagues for their effort and support.

\end{document}